\documentclass{article}

\usepackage{arxiv}

\usepackage[utf8]{inputenc} 
\usepackage[T1]{fontenc}    
\usepackage[hidelinks]{hyperref}       
\usepackage{url}            
\usepackage{booktabs}       
\usepackage{amsfonts}       
\usepackage{nicefrac}       
\usepackage{microtype}      
\usepackage{graphicx}
\usepackage{doi}
\usepackage{xcolor}
\usepackage{enumitem}
\usepackage{makecell}
\usepackage[round, authoryear]{natbib}

\PassOptionsToPackage{hyphens}{url}
\usepackage[x11names]{xcolor}
\IfFileExists{xurl.sty}{\usepackage{xurl}}{} 
\hypersetup{
  pdftitle={Insights for an AI Whistleblower Office From 30 Case Studies},
  colorlinks=true,linkcolor=black,urlcolor=RoyalBlue4,citecolor=RoyalBlue4}

\title{
Insights for an AI Whistleblower Office \\ 
From 30 Case Studies
}

\author{ 
        {\hspace{1mm}Ethan Beri}\thanks{Correspondence to \texttt{ethan.beri@pmb.ox.ac.uk}. \\ Ethan Beri designed the research methodology, constructed and analysed the dataset, and wrote the report. Mauricio Baker supervised the project and gave feedback throughout. This paper represents only the authors' views, and not those of the University of Oxford or RAND.} \\
	University of Oxford \\
	\And
	{\hspace{1mm}Mauricio Baker} \\
	RAND  \\
}

\date{ }

\renewcommand{\headeright}{ }
\renewcommand{\undertitle}{ }
\renewcommand{\shorttitle}{ }

\hypersetup{
pdftitle={Insights for an AI Whistleblower Office From 30 Case Studies},
pdfauthor={Ethan Beri, Mauricio Baker},
}

\begin{document}
\maketitle

\begin{abstract}
    Whistleblower programmes are a promising tool for uncovering noncompliance with AI regulations. This paper aims to help policymakers design an AI whistleblower programme by giving them an understanding of whistleblowers’ motivations, and of the overall whistleblowing process. We take an empirical approach, assembling a dataset of 30 case studies of whistleblowers. This dataset includes dozens of features of each case, which range from 1978 to 2020 and span 15 industries. Our findings suggest that whistleblower programmes will be more effective if they financially reward whistleblowers, provide protections for whistleblowers, enable whistleblowers to report anonymously, are adequately staffed and funded, and provide advice to potential whistleblowers. We provide ten concrete policy recommendations for an AI whistleblower programme at the end of this paper.
\end{abstract}

\section{Executive summary}

Whistleblower programmes play a significant role in uncovering wrongdoing in a number of industries. For example, between 2010 and 2023, the SEC’s whistleblower programme recovered more than \$4 billion in penalties \citep{secpress2023}. Whistleblower programmes have been “praised by the SEC and CFTC Chairs, other Republican and Democratic political leaders, … academics, and many others” \citep{platt2023}. \textbf{Given their strong track record and bipartisan support, whistleblower programmes could also be a valuable tool for regulating AI.} We define a “whistleblower programme” as a centralised, industry-wide office designed to receive, process, and investigate “tips” (information about wrongdoing) and provide incentives to whistleblowers (such as financial payments or protection). 

\textbf{To effectively design a whistleblower programme, policymakers need an understanding of the whistleblowing process, and of whistleblowers’ motivations.} However, there is limited public data available. \textbf{To fill this gap, we assembled a dataset of 30 case studies of whistleblowers,} selected as a random sample from public lists of notable cases. Our dataset compiles open-source information about dozens of aspects of these cases, which range from 1978 to 2020 and span 15 industries.

\textbf{Our dataset shows a number of common themes.} The typical whistleblower in our sample was middle-aged, well-educated, and doing well for themselves within a large organisation. Over 90\% were insiders, of which roughly 80\% were mid-level employees or executives. Upon discovering wrongdoing, at least 87\% of whistleblowers reported it for apparently moral reasons\footnote{See our methodology and findings sections for more information on how we attributed motivation to different cases. We argue that this surprising result is not explained by biases involved in reporting cases. It is also worth noting that whistleblowers often had multiple motivations, and were not solely morally motivated.}. In reporting wrongdoing, whistleblowers often had to overcome their fears of retaliation and social pressure, as well as their lack of trust in action being taken on the basis of their report (each present in 40-80\% of cases). Usually, they did not report anonymously—just 13\% of whistleblowers did so—and tended to move towards more independent, external actors as the whistleblowing process went on. Most of the time, they were retaliated against (57-67\% of cases). Often, whistleblowing harmed their career prospects (43-57\% of cases). For an unlucky few, however, retaliation threatened their lives and freedoms: four whistleblowers (13\%) received death threats, including three (10\%) who were employees of private companies. 

\textbf{This data suggests that an effective AI whistleblower programme should have the following features:}

\begin{enumerate}[series=execsumm]
    \item \textbf{Financial rewards.} Financial incentives played a moderate role in our dataset (27\% of cases), despite rarely being available. Whistleblower rewards also have broad support in the microeconomics and behavioural economics literature. This suggests that AI whistleblowers should receive a percentage of the sanction collected. The exact percentage could be similar to the SEC and CFTC programmes, which award whistleblowers 10-30\% of the amount of the sanction resulting from their tip. 
    \item \textbf{Protection.} 57-67\% of whistleblowers in our sample were retaliated against, and the fear of retaliation was a demotivating factor in over 40\% of cases. Protecting whistleblowers could make them significantly more likely to report wrongdoing. Options for policymakers include a) prohibiting employers from retaliating against whistleblowers (as in Dodd-Frank Section 922) \citep{doddf922}, and b) placing whistleblowers in witness protection programmes (which would be especially useful for cases such as the 13\% of our dataset which received death threats). Additionally, officials could c) grant S visas to international AI whistleblowers.
    \item \textbf{Anonymity.} As another form of protection from retaliation, enabling whistleblowers to tip anonymously may make them more likely to report wrongdoing. AI whistleblowers should be able to tip anonymously with the help of a lawyer, or through an anonymous online platform. The AI whistleblower office could also take cybersecurity measures to protect whistleblowers' identities.
    \item \textbf{Processing tips.} Processing tips effectively is important for attracting future tips. Adequate staffing and funding is necessary for effective "tip-sifting". 
    \item \textbf{Messaging and advice.} Potential whistleblowers need to be aware of the programme, and need to trust it enough to tip. Previously, other whistleblower programmes have had slow starts, possibly due to a lack of messaging or awareness. Mandatory trainings and public endorsements could build trust and awareness for the programme. Additionally, the AI whistleblower office could offer advice to prospective whistleblowers experiencing psychological distress, or who are unsure that they have reasonable cause to believe that violations have occurred. 
\end{enumerate}

\href{https://github.com/ethandberi/whistleblowers}{\textcolor{black}{Our dataset is publicly available} \textcolor{blue}{here}.}

\begin{figure}[!h]
  \centering
  \begin{minipage}[b]{0.48\textwidth}
    \raisebox{8.3pt}{\includegraphics[width=\textwidth]{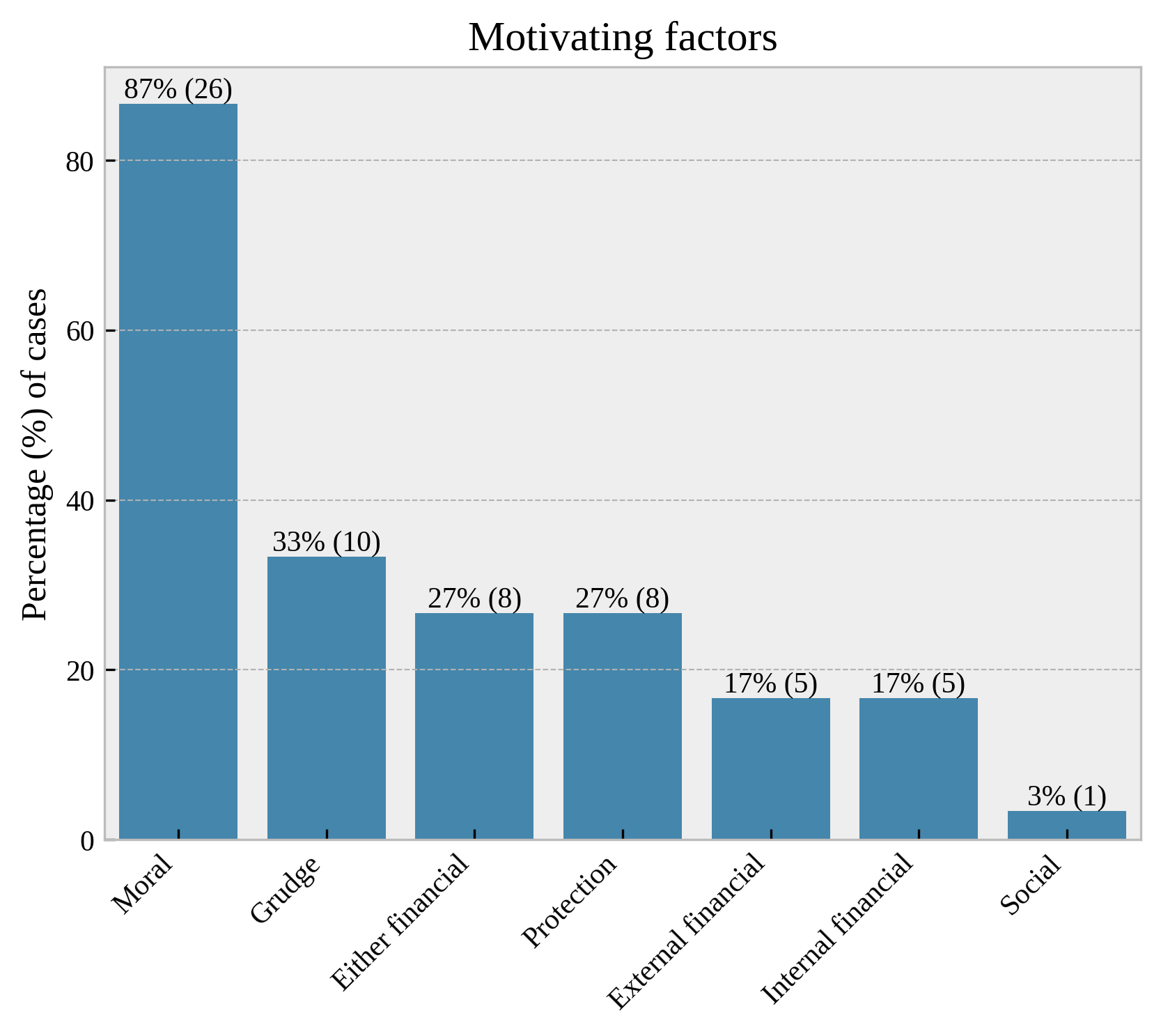}}
  \end{minipage}
  \hfill
  \begin{minipage}[b]{0.48\textwidth}
    \includegraphics[width=\textwidth]{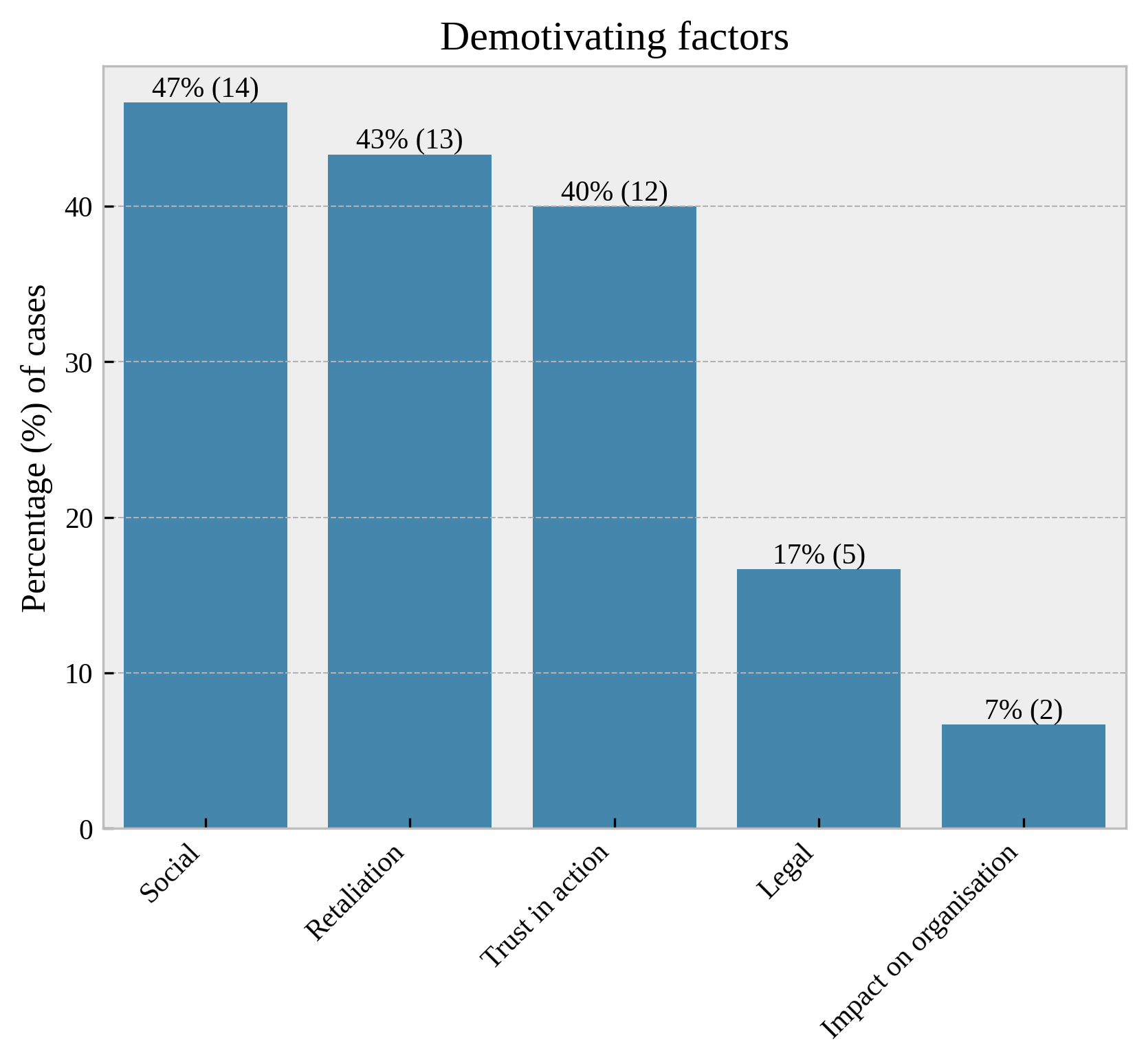}
  \end{minipage}
  \caption{Prevalence of motivating (left) and demotivating (right) factors. Note we took a conservative, careful approach to ascribing motivation: each case used a range of sources, and we preferred a sparse field to an incorrect one. More on this in Methodology and dataset.}
\end{figure}

\newpage

\tableofcontents

\newpage

\section{Background}

\textbf{We define a “whistleblower programme” as a centralised office designed to receive, process, and investigate “tips” (information about wrongdoing) and provide incentives to whistleblowers (such as financial payments or protection).} In this paper, we focus on industry-wide whistleblower programmes, as opposed to programmes operating within a single company, or across multiple industries. Specifically, we aim to help policymakers design a whistleblower programme for the AI industry.

\textbf{Whistleblower programmes could be a force multiplier in AI regulation:} by helping to uncover noncompliance, whistleblower programmes could make a host of other regulations more effective. \textbf{Examples of AI regulations which could benefit from whistleblower programmes include:}

\begin{itemize}
    \item \textbf{Domestic AI regulation:} a whistleblower programme for domestic AI regulations could play a similar role to those at the SEC and CFTC in finance.
    \item \textbf{Export controls:} since the introduction of export controls on AI chips in 2022, their illegal smuggling into China has become a national security issue \citep{nyt2024, wsj2024, ft2024}. A whistleblower programme could help bring cases to the attention of the Bureau of Industry and Security \citep{cnas2025}.
    \item \textbf{International agreements:} proposed international agreements on AI aim to avoid “race to the bottom” dynamics between countries. However, without sufficient risk of being discovered, countries may have an incentive to defect from these agreements: they could reap the benefits of others agreeing (for example, by slowing their rivals’ AI development efforts), without incurring any cost to themselves. An international whistleblower programme could make defection a less attractive option by increasing the likelihood of defectors being discovered \citep{baker2025, hazza2025}. 
\end{itemize}

\textbf{Whistleblower programmes must solve two problems: attracting tips, and processing them. We focus exclusively on the former problem.} What might encourage whistleblowers to come forward? What might deter them? These questions are best answered empirically, though little research of this kind has been published. We know of only one other paper dedicated to designing whistleblower policy for AI, namely \cite{wu2025}, which takes a more theoretical approach. Indeed, only a few empirical studies of whistleblowing exist at all \citep{tnejom2010, vandy2019}. A handful of papers in microeconomics attempt to design optimal whistleblower programmes, but say little about whistleblowers’ motivations \citep{givat2016, givat2017}. This is a significant limitation: it has been shown that optimal policy depends on whistleblowers’ motives \citep{heyeskapur2009}\footnote{ Note that this is not a complete literature review; we only attempt to sketch the research landscape. For more information on the whistleblower literature, readers should contact the corresponding author.}. In short: \textbf{to design a whistleblower programme for AI regulation, policymakers need an empirical understanding of the whistleblowing process and whistleblower motivation.}

\section{Methodology and dataset}

Studying whistleblowers can produce a rich picture of their motivation. However, there is limited public data available on whistleblowers’ motivations and the whistleblowing process. This may be because whistleblowers sometimes choose to report anonymously, and because programmes like the SEC’s and CFTC’s strive to protect whistleblowers' identities. 

\textbf{To address this lack of data, we assembled a dataset of 30 case studies, randomly sampled from open-source lists of notable whistleblowers.} Specifically, these lists are sourced from the Government Accountability Project and Wikipedia, and only cases from 1970 onwards \citep{gaplist, wikipedialist}. Note that Wikipedia was only used for its list of whistleblowers, and not as a source of information. No claims in our data or this article depend on the truthfulness of Wikipedia articles. 

\textbf{For each case, a number of articles from reputable sources were used to fill 58 fields}. Regarding sources, we prioritised using a combination of first-hand interviews (ideally with the whistleblower) and articles published by reputable news outlets, such as The Washington Post, The New York Times, The Guardian, and others. Where applicable, we also used peer-reviewed academic sources.

Fields included questions about 1) the whistleblower, 2) the organisation responsible for the wrongdoing, 3) the act of wrongdoing, 4) the act of whistleblowing, 5) the whistleblower-organisation relationship (for example, whether the whistleblower was an insider), 6) the whistleblower-wrongdoing relationship (for example, whether the whistleblower was complicit), and 7) the whistleblower’s motivation. Typically, the time spent researching a single case ranged from 90-180 minutes.

\textbf{In general, we took a conservative approach in assembling this dataset: we prefer a sparse field to an incorrect one.} This was especially relevant when attributing motivation to whistleblowers. In particular, we attempted to avoid reporting (and self-reporting) bias: that is, when reporters or whistleblowers have an incentive to misrepresent their motives. For example, whistleblowers have an incentive to report altruistic motives, because it is better for their reputation than if they were to report more “selfish” motives. We accounted for this by considering multiple reported points of view, and by comparing the whistleblowers’ incentives to those of other parties (for example, potential whistleblowers who chose not to tip). To help readers and dataset users fully understand our reasoning for these fields, we have included an evidence column after each motivation column in the dataset.

Note that we also created (but did not release) other evidence fields for more difficult or judgement-based topics. We also did not release a number of other fields, usually because they were too sparse. This data is available on request.

\section{Limitations}

Our method of sampling—that is, randomly selecting 30 cases from open-source lists—comes with three main limitations. The first is that \textbf{our sample only includes cases where whistleblowers tipped successfully} (an example of “selecting for the dependent variable” \citep{geddes1990}). Ideally, our dataset would contain all of 1) cases of successful tipping (as it does now), 2) cases of unsuccessful tipping (that is, tipping with no effect on wrongdoing), and 3) cases where wrongdoing was not reported in the first place. This would make it easier to claim that certain variables about each case made the difference between tipping and not tipping, or between success and failure. 

Unfortunately, cases like (2) or (3) are often not made public. To mitigate the absence of category (2), we made an effort to think about whistleblowing as a protracted process, often involving many rounds of unsuccessful tipping\footnote{For more on this model of the whistleblowing process, see \cite{vandy2019}. Based on our own data, we believe that the “protracted process” model is accurate.}. For example, fields in our dataset compare the initial and final steps in the whistleblowing process. Additionally, to make up for the lack of cases like (3), we made a conscious effort to record data on potential whistleblowers who chose not to report (such as whistleblowers’ colleagues)\footnote{ Note that fields solely concerning whistleblowers’ colleagues were not released—on the basis of being too imprecise and too sparse—but this data was used in our reasoning in other fields (such as when attributing motivation). Therefore, for cases in which these facts were important, they can be found in our evidence columns.}. Comparing these two groups’ circumstances—those who blew the whistle, and those who chose not to—should paint a more accurate picture of each case, and of what motivates whistleblowing.

\textbf{Our sample is also biased towards famous cases.} This affects the external validity of our results: our sample may not accurately represent all whistleblowers, because we only sampled from lists of well-known cases. For example, the small proportion of whistleblowers in our sample who sought anonymity—just 13\%—is unlikely to be the same for the general whistleblower population. Instead, this result should probably be attributed to sampling bias: we only used well-known cases, and well-known cases may be less likely to involve an anonymous whistleblower (perhaps because knowing the whistleblower’s identity makes for a better news story). 

However, our sample may not be so biased in other data fields. Some patterns from our data match the findings from other empirical studies of whistleblowers—studies which do not suffer from our sample’s bias towards famous cases—which may suggest that our dataset still reflects important facts about the whistleblower population. One of these patterns is that whistleblowers tend to move towards more independent, external methods of reporting with each step in the whistleblowing process \citep{vandy2019}. This is mirrored in our dataset: only 50\% of whistleblowers’ first tips were to an external party, which climbs to over 80\% for their last tips. Another pattern is that whistleblowers are often motivated (at least in part) by moral considerations: one study of pharmaceutical whistleblowers found that their top three motivations were integrity, justice, and altruism \citep{tnejom2010}\footnote{Note that this study consisted of individuals who had filed \textit{qui tam} lawsuits under the Federal Claims Act, which entitles them to receive 15-25\% of financial recoveries from their case. This makes their finding even more striking: even where substantial financial rewards were available, moral motivations still played a significant role in decision-making.
}. This is compatible with our results: at least 87\% of our sample’s whistleblowers appeared to have some moral motivation. In summary: while our data likely suffers from sampling bias in some areas (such as anonymity), it can still be useful overall. 

\textbf{Finally, our data uses a relatively small sample (n=30).} Again, this makes our data less likely to be representative of the relevant whistleblower population. Similar points can be made to justify its usefulness in spite of this limitation. Pragmatically, our sample was small because we prioritised having a high-dimensional dataset over a high-n one: we were more interested in deeply understanding a few cases than in having a shallow understanding of many. 

\section{Findings}

In this section, we summarise trends in our dataset. Due to limited data availability in some cases, we often present ranges or bounds rather than exact values. For example, although 15/30 whistleblowers have a postgraduate degree, there are 5 more whose education is unknown. So the true percentage for the sample is at least 50\% (assuming all unknown cases \textit{do not} have a postgraduate degree), and at most 67\% (assuming all unknown cases \textit{do} have a postgraduate degree). Therefore, we present the proportion of whistleblowers with a postgraduate degree as being between 50-67\%. Similar reasoning is used wherever percentage ranges are present.

\subsection{Demographics}

Whistleblowers in our dataset tended to be \textbf{male} (70-73\%), \textbf{middle-aged} (average age of 45, median of 44)\footnote{These figures are calculated from exact ages, available for 23/30 cases.}, and \textbf{well-educated}, with 50-67\% having a postgraduate degree. 

\begin{figure}[!h]
  \centering
  \begin{minipage}[t]{0.48\textwidth}
    \includegraphics[width=\textwidth]{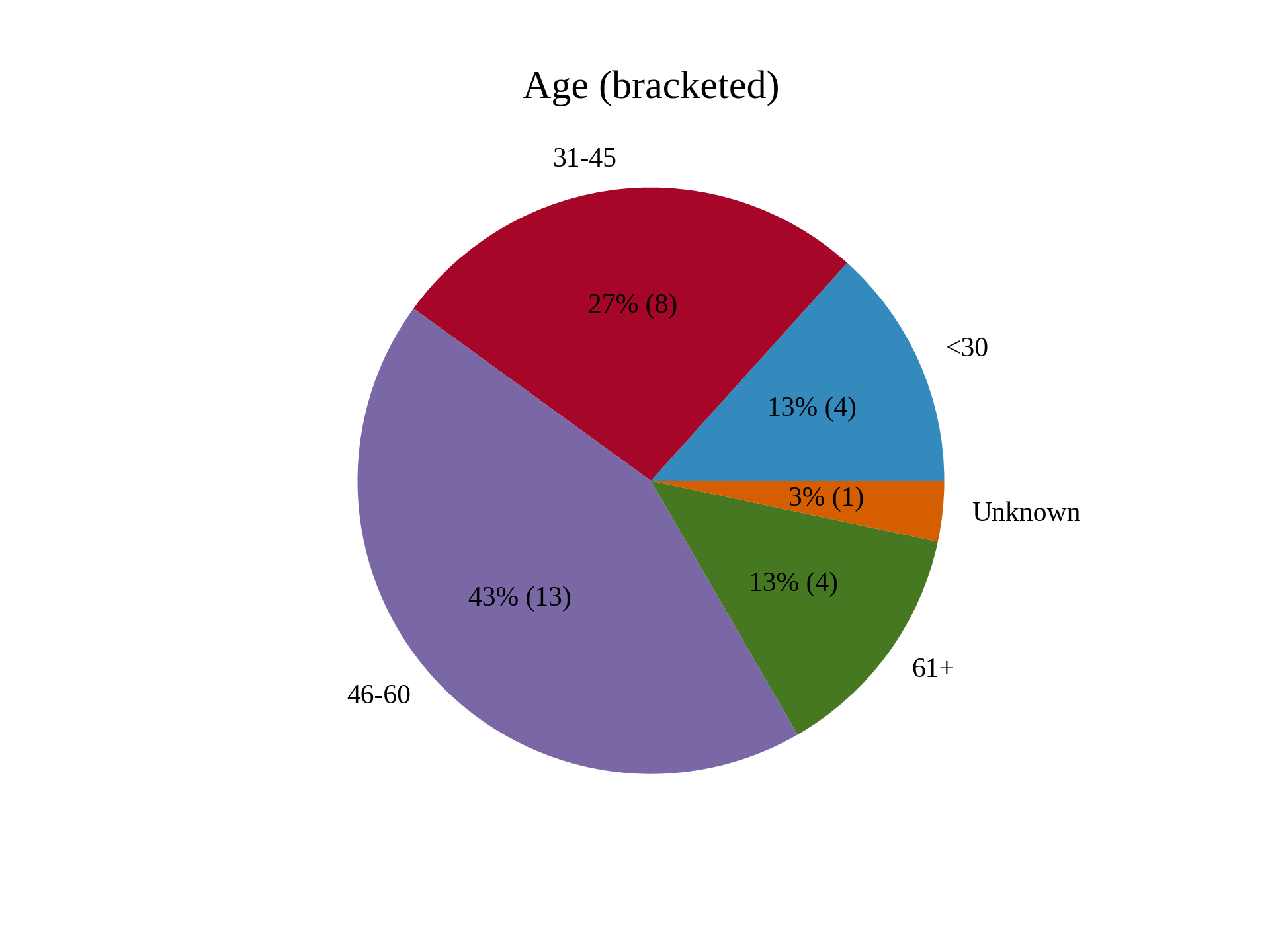}
    \caption{Whistleblower ages at the time of making their first report, bracketed (to include estimated ages, in cases where exact ages were unavailable).}
  \end{minipage}
  \hfill
  \begin{minipage}[t]{0.48\textwidth}
    \includegraphics[width=\textwidth]{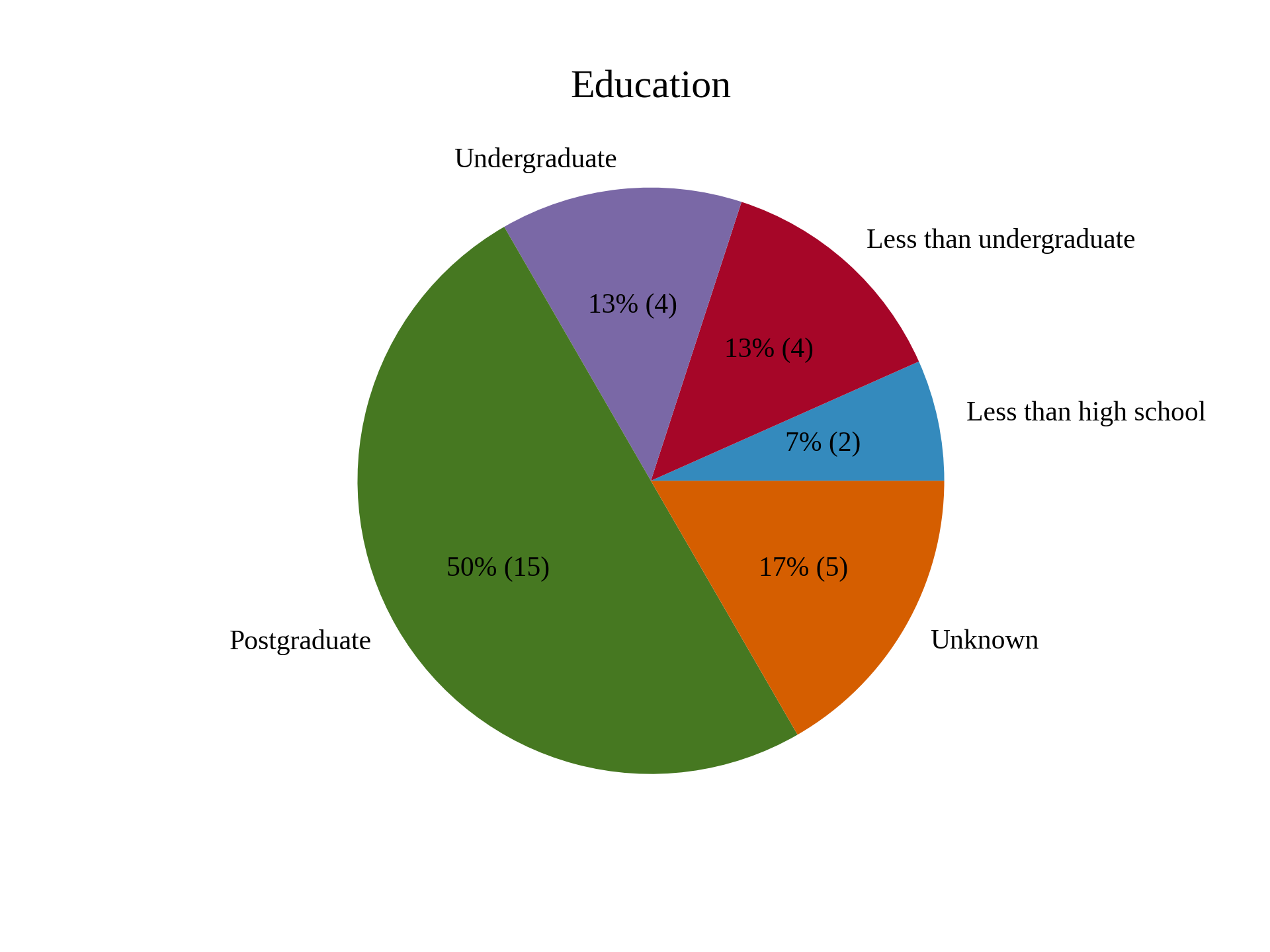}
    \caption{Whistleblower education levels.}
  \end{minipage}
\end{figure}

\subsection{Organisations}

\textbf{Cases mostly involved large organisations}, with over 10,000 employees (57-73\%)\footnote{Note this is calculated per case, not per organisation: there is one organisation (the NSA) which appears three times in the dataset. 
}. Using a combination of exact figures and estimates, the median organisation had 38,425 employees\footnote{This number comes from combining the 9 cases where exact employee numbers are available with the 16 cases where it was feasible to estimate this number. Estimates fell into a range (1-50, 51-100, 101-500, 501-1,000, 1,001-5,000, 5,001-10,000, 10,001-100,000, 100,001+) which were then interpolated (for example, 1-50 would become 26) and used as if they were exact figures. Note that 100,001+ was treated as 100,001.}. Excluding cases where exact figures are not available, the median size comes down to 26,921. 

\textbf{There was no “typical” sector or industry.} Sectors were split between government (43\%) and for-profit (53\%), with one not-for-profit. Industries were varied, with the most common being intelligence (20\%) and energy (17\%). 

\textbf{Most cases involved organisations which had been the subject of whistleblowing before} (at least 67\%) \textbf{and would be again in the future} (at least 53\%)\footnote{Note the use of “at least”: we may not have found all occasions when these organisations were the subject of whistleblowing.}. Few organisations had never been the subject of a whistleblower case before or after their case in the sample (at most 10\%).

\textbf{Almost all whistleblowers were employees of the organisation which performed the wrongdoing} (at least 90\%)\footnote{Insiders comprise at most 93\% of our sample—it is unclear whether the Panama Papers whistleblower was an insider. The two certain non-insiders are Harry Markopolos and Erin Brockovich.}. \textbf{Of these insiders, roughly 80\% were mid-level employees or executives.}

\begin{figure}[!h]
  \centering
  \begin{minipage}[t]{0.48\textwidth}
    \includegraphics[width=\textwidth]{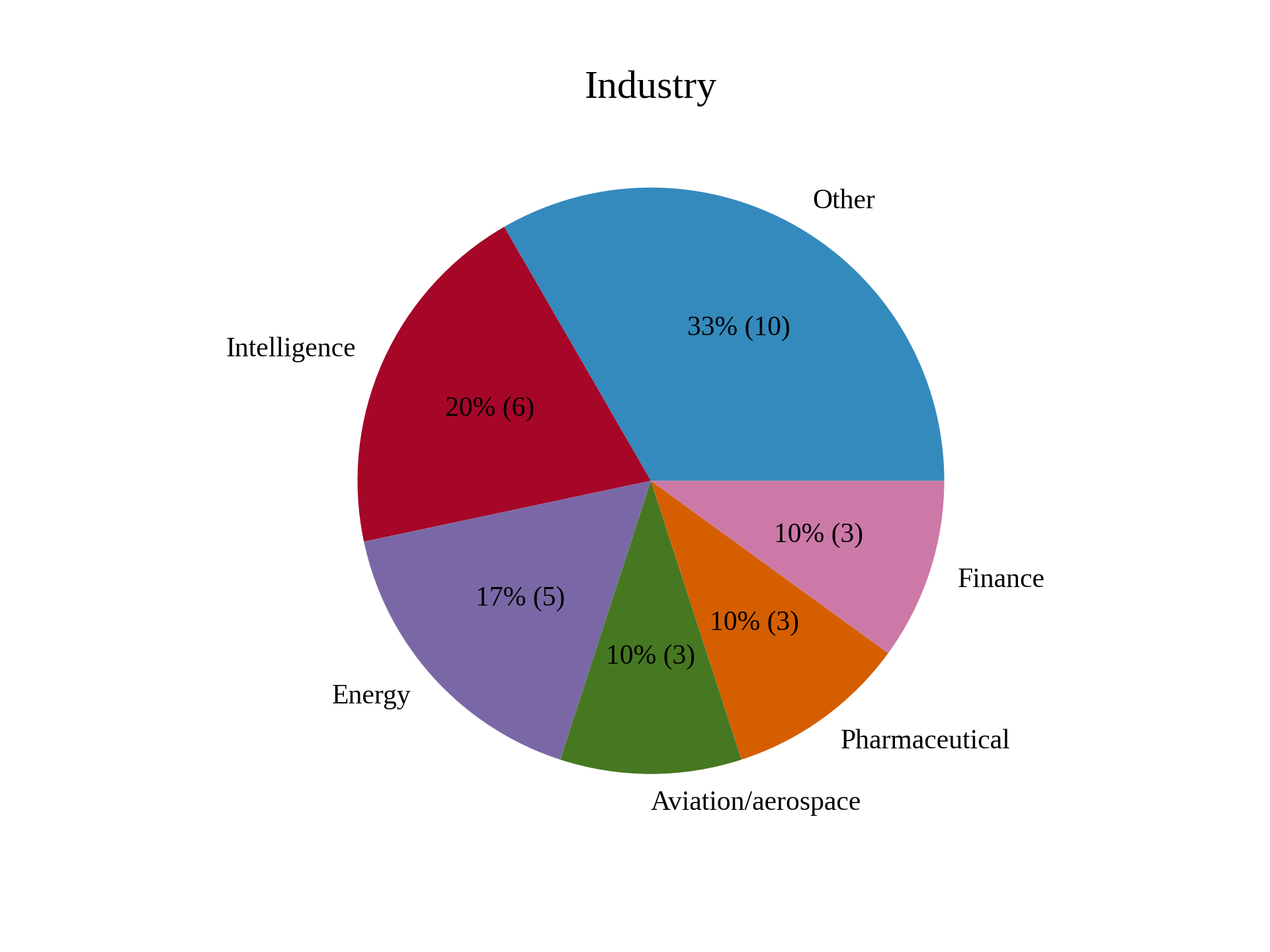}
    \caption{Wrongdoer organisations' industries.}
  \end{minipage}
  \hfill
  \begin{minipage}[t]{0.48\textwidth}
    \includegraphics[width=\textwidth]{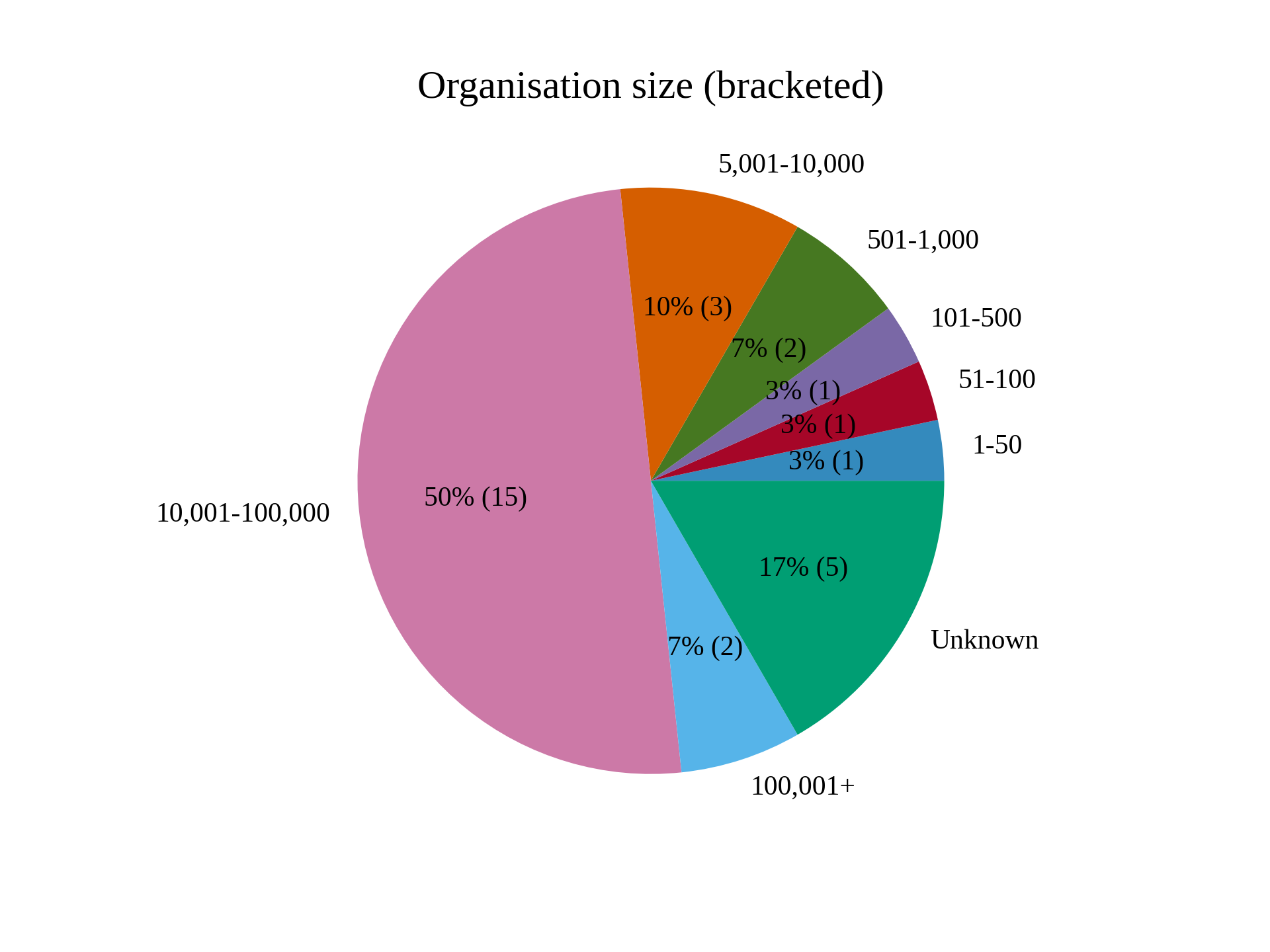}
    \caption{Sizes of wrongdoer organisations.}
  \end{minipage}
\end{figure}


\subsection{Wrongdoing}

\textbf{Types of wrongdoing were varied}, the most common being fraud (23\%) and safety violations (20\%).

\textbf{In the vast majority of cases, wrongdoing was ongoing} (80\%), rather than a one-off event. Of the 19 cases both 1) involving ongoing wrongdoing and 2) with a known end date, \textbf{wrongdoing generally lasted around 3 years}\footnote{Median 3 years; mean approximately 7 years; upper quartile 8 years. There were two notable outliers (26 and 38 years).
}.

\textbf{Most cases in our dataset occurred in the U.S.} (73\%), likely due to sampling bias: for example, the Government Accountability Project—which created one of the lists of notable whistleblowers used in our sample—is based in Washington, D.C. and seemingly focuses on working with U.S. whistleblowers. 

\textbf{Some whistleblowers were complicit} (20-33\%), \textbf{and almost all had complete knowledge of the wrongdoing} (at least 90\%)\footnote{“Complete” knowledge does not imply that the whistleblower knew everything about the wrongdoing (for example, the exact dollar amount of a fraud). Instead, a whistleblower has “complete” knowledge of wrongdoing if they are aware of its general scope and nature (for example, knowing that mass fraud was being committed).}. 

\begin{figure}[!h]
    \centering
    \includegraphics[width=0.48\linewidth]{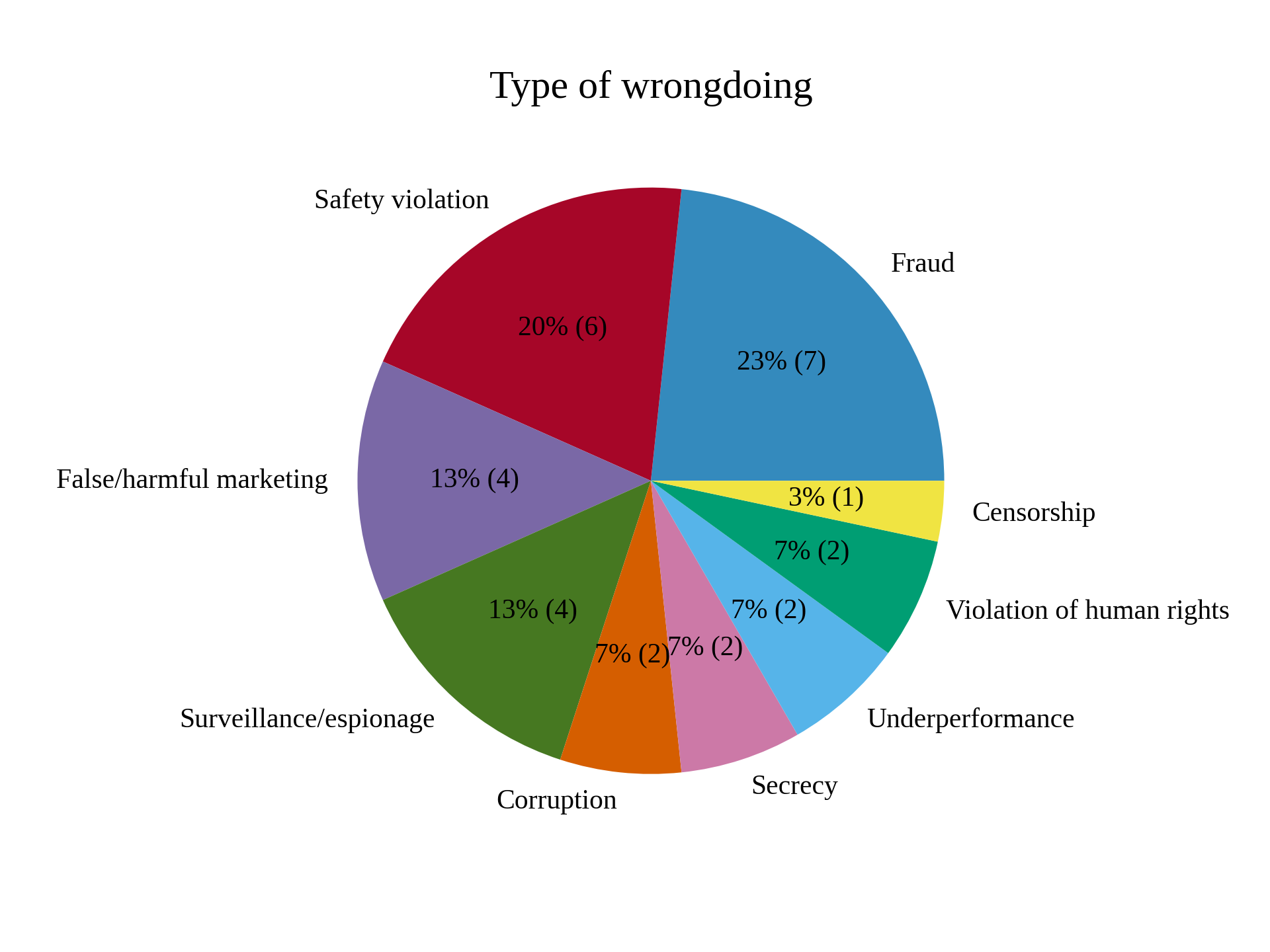}
    \caption{Types of wrongdoing committed. Note the significant variation.}
\end{figure}

\subsection{Whistleblowing process}

\textbf{Most whistleblowers tipped (internally or externally) within a year of discovering wrongdoing:} at least 67\% made their first tip a year or less after discovering the wrongdoing. Of the 25 cases where data on this field was available, the mean time from discovery to initial tip was less than a year.

\textbf{As the whistleblowing process went on, whistleblowers moved towards increasingly external parties.} Whistleblowers’ initial tips are split between internal (50\%; 15/30) and external (50\%; 15/30) parties, but their final tips are overwhelmingly to external parties (at least 83\%). 

\textbf{Of internal whistleblowers} (at least 90\% of the sample), \textbf{most tipped during their employment} (at least 77\%), \textbf{not after.}

\textbf{In approximately half of all cases, whistleblowing had a negative impact on the whistleblower’s career} (43-57\%). For a few, whistleblowing had a positive impact (17-30\%), and for roughly a third of cases, there was little or no impact (23-40\%)\footnote{ It may be hard to imagine how whistleblowing could have a positive impact on an individual’s career. As an example, consider Erin Brockovich, who famously blew the whistle on Pacific Gas and Electric: “Brockovich is the first to admit that the case changed her life. For a start, she received \$2.5m (£1.73m) as part of her fee from the settlement, which meant she was able to move out of a cockroach-infested dive into a mansion in Agoura Hills, in the San Fernando Valley suburbs of Los Angeles …” \citep{tga2001}. Brockovich now works as a clean water activist, helped by her whistleblower credentials \citep{tatl2020}.}. 

\subsection{Anonymity}

\textbf{Very few whistleblowers sought anonymity:} just 4 known cases, which is 13\% of the sample\footnote{ We specify \textit{known} cases because it is possible that other whistleblowers initially sought anonymity, which may not have been reported in the sources we used. However, because we used a variety of sources for each case and focused on building comprehensive narratives, we believe that the omission of an important detail like this is unlikely.}. Out of those who did seek anonymity, all received it (100\%; 4/4)\footnote{We use “received” for lack of a better term: these whistleblowers were either kept anonymous by external parties, or successfully kept themselves anonymous (such as the Panama Papers whistleblower).}. These cases include Tamm, whose identity was kept anonymous by New York Times reporters \citep{pbs2013}; the Panama Papers whistleblower, who kept their own identity anonymous \citep{spieg2022}; Manning, who anonymously leaked to WikiLeaks \citep{guard2022}; and Winner, who was briefly anonymous when she leaked to The Intercept \citep{howl2017}. \textbf{Of these four, only Manning and Winner had their anonymity broken (50\%; 2/4)\footnote{ Note that we do not count Tamm’s anonymity as having been “broken”, because he told the New York Times reporters that they “did not have to go to jail for [him]”, and it seems likely that they were the ones to inform the FBI about Tamm \citep{pbs2013}.}.}

\subsection{Retaliation}

\textbf{Most whistleblowers were retaliated against} (57-67\%). \textbf{The most common forms of retaliation were harassment}\footnote{“Harassment” was used broadly to mean action taken with the purpose of inflicting psychological harm. This could range from workplace hostility to anything short of death threats.} (27-37\%) \textbf{and unjust termination} (20-30\%). A significant percentage of whistleblowers received death threats (13-23\%)\footnote{The 4/30 whistleblowers who received death threats were: John Michael Gavitt (who blew the whistle on GE Aircraft Engines), Arnold Gundersen (Nuclear Energy Services), Cristoph Meili (Union Bank of Switzerland), and Mukesh Kapila (United Nations, Sudanese government).}.

\begin{figure}[!h]
    \centering
    \includegraphics[width=0.6\linewidth]{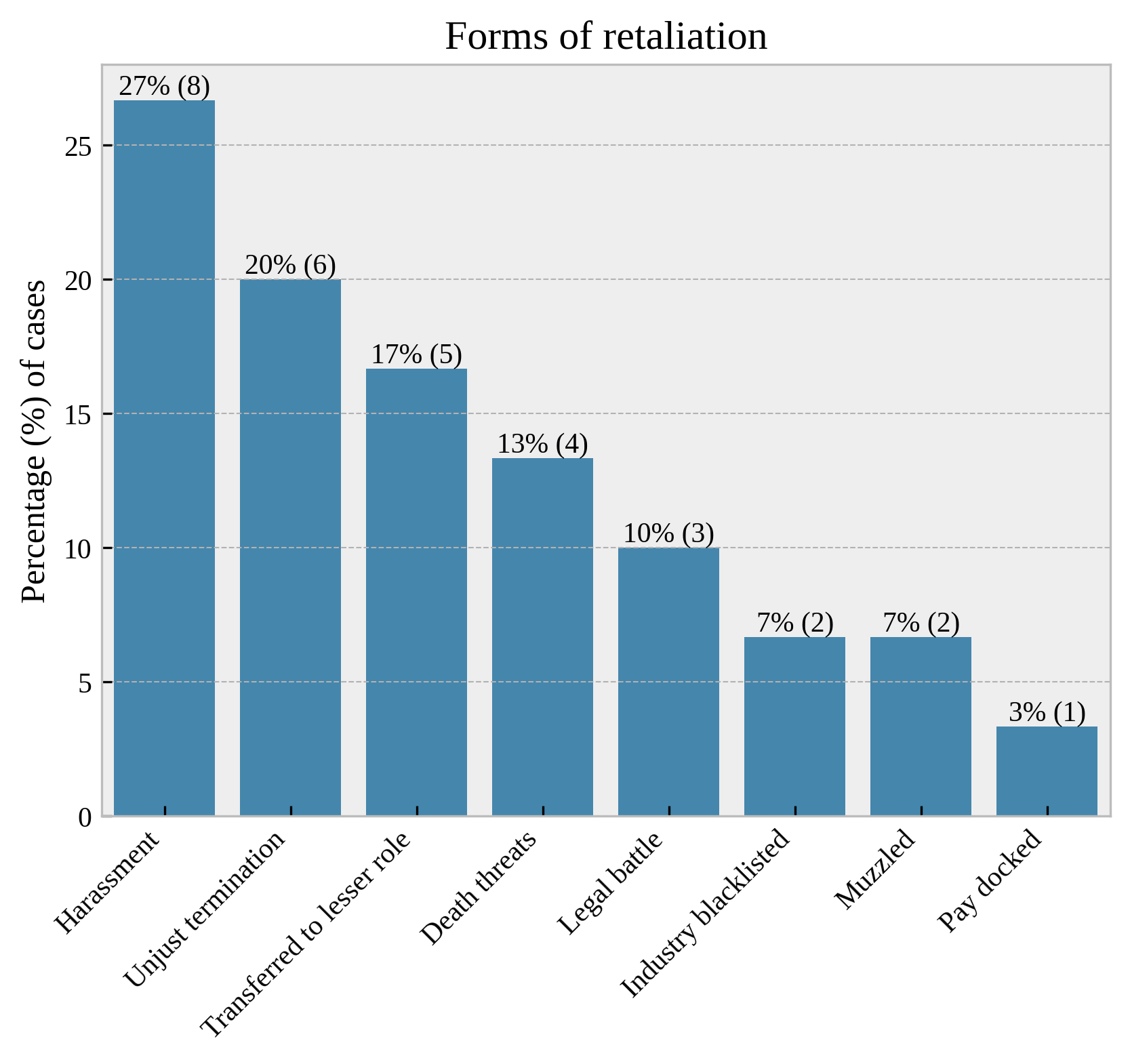}
    \caption{Forms of retaliation. Percentage labels are calculated from the size of the entire sample (30), not from the subset who were retaliated against (at least 17), hence all percentages are lower bounds. Note that “Muzzled” here means “Directed by higher-ups to stop speaking publicly about the case”.}
\end{figure}

\subsection{Motivation}

\textbf{Almost all whistleblowers appeared to be (at least partly) morally motivated} (at least 87\%). Some whistleblowers appeared \textit{solely} morally motivated, with no other apparent source of motivation (at least 7\%)\footnote{The proportion of solely morally motivated whistleblowers is 7-37\%. This range is so large because many cases have at least one kind of incentive for which no data is available: therefore, assuming that the unavailable incentive would have been present significantly decreases the number of solely morally motivated cases, and vice versa.}. \textbf{Some also appeared to be motivated by grudges} (at least 33\%), though there is less data available here. 

\begin{figure}[!h]
    \centering
    \includegraphics[width=0.6\linewidth]{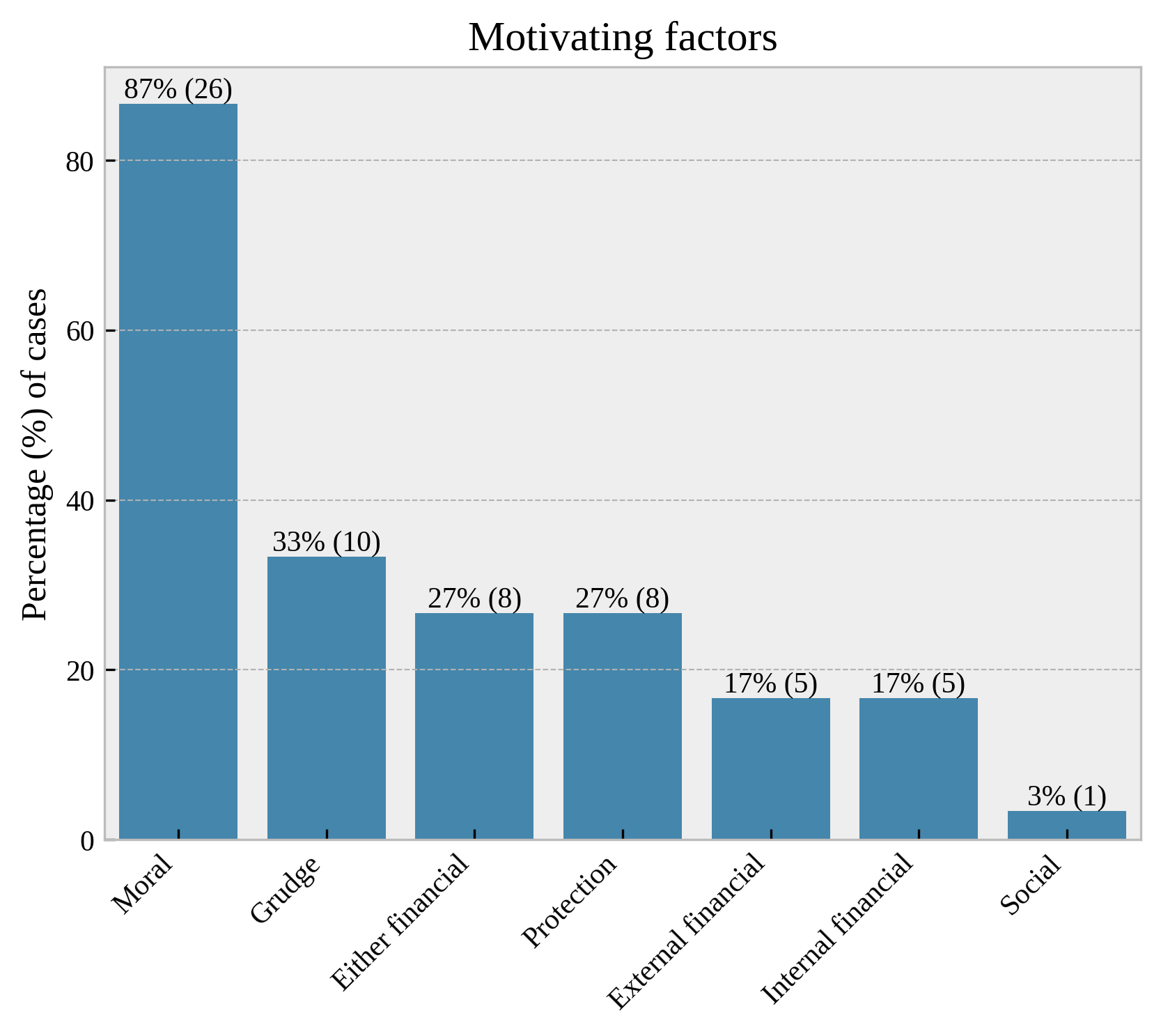}
    \caption{Motivating factors present for whistleblowers. All percentages are lower bounds.}
\end{figure}

\begin{table}[!h]
    \caption{Motivating factors, including "Unclear" cases.}
    \centering
    \begin{tabular}{cccccccc}
        \toprule
        & Moral & \makecell{Financial\\(external)} & \makecell{Financial\\(internal)} & \makecell{Financial\\(either)} & Social & Grudge & Protection/anonymity \\
        \midrule
        Yes & 26 (87\%) & 5 (17\%) & 5 (17\%) & 8 (27\%) & 1 (3\%) & 10 (33\%) & 8 (27\%) \\
        No & 0 (0\%) & 25 (83\%) & 24 (80\%) & 22 (73\%) & 19 (63\%) & 11 (37\%) & 19 (63\%) \\
        Unclear & 4 (13\%) & 0 (0\%) & 1 (3\%) & 0 (0\%) & 10 (33\%) & 9 (30\%) & 3 (10\%) \\
        \bottomrule
    \end{tabular}
\end{table}

See footnotes for the definitions of external financial motives\footnote{"External” financial motives are offered by external parties (i.e. not the organisation responsible for the wrongdoing, or—for non-insiders—the whistleblower’s own organisation). For example, the prospect of receiving a financial reward from filing a False Claims Act suit under the \textit{qui tam} provision would count as an external financial motive. (Note the \textit{prospect} of a reward, not an \textit{actual} reward, is what counts here: in motivating whether a whistleblower submits a tip or not, what matters is the expected, not actual, payoff).}, internal financial motives\footnote{An “internal” financial motive is offered by the whistleblower’s own organisation. For example, if an individual is working as an auditor and uncovers wrongdoing within their company, they can plausibly expect some monetary payoff (an internal financial motive) in the future.}, and social motives\footnote{A “social” motive would be present if a whistleblower expects to gain social status from whistleblowing (such as fame, or a reputation for being an upstanding citizen).}. We also explain how eight whistleblowers were motivated by protection or anonymity, despite only four having actually sought anonymity\footnote{Although only four whistleblowers sought anonymity, eight were motivated by anonymity or protection. This eight includes all four anonymity-seeking whistleblowers, as well as four others who sought protection or anonymity at some point in the whistleblowing process. These four others were David Lochridge, Rick Bright, John Barnett, and Walter Tamosaitis. Field \texttt{protection\_incentive\_notes} of our dataset explains their protection/anonymity motives further.}. 

\textbf{In most cases, the main demotivating factors present were retaliation} (43-83\%), \textbf{a lack of trust in action being taken on the basis of a tip} (40-83\%), \textbf{and social factors} (47-77\%)\footnote{Note that these percentage ranges are so large because these fields were much more sparse. This is because it was much more difficult to identify what demotivating factors a whistleblower had faced (and yet overcome) than it was to identify what had motivated them to overcome those factors in the first place. For example, take the lack of trust in action being taken: sufficiently confident answers could only be given for 17 cases for this field.}. 

\begin{figure}[!h]
    \centering
    \includegraphics[width=0.6\linewidth]{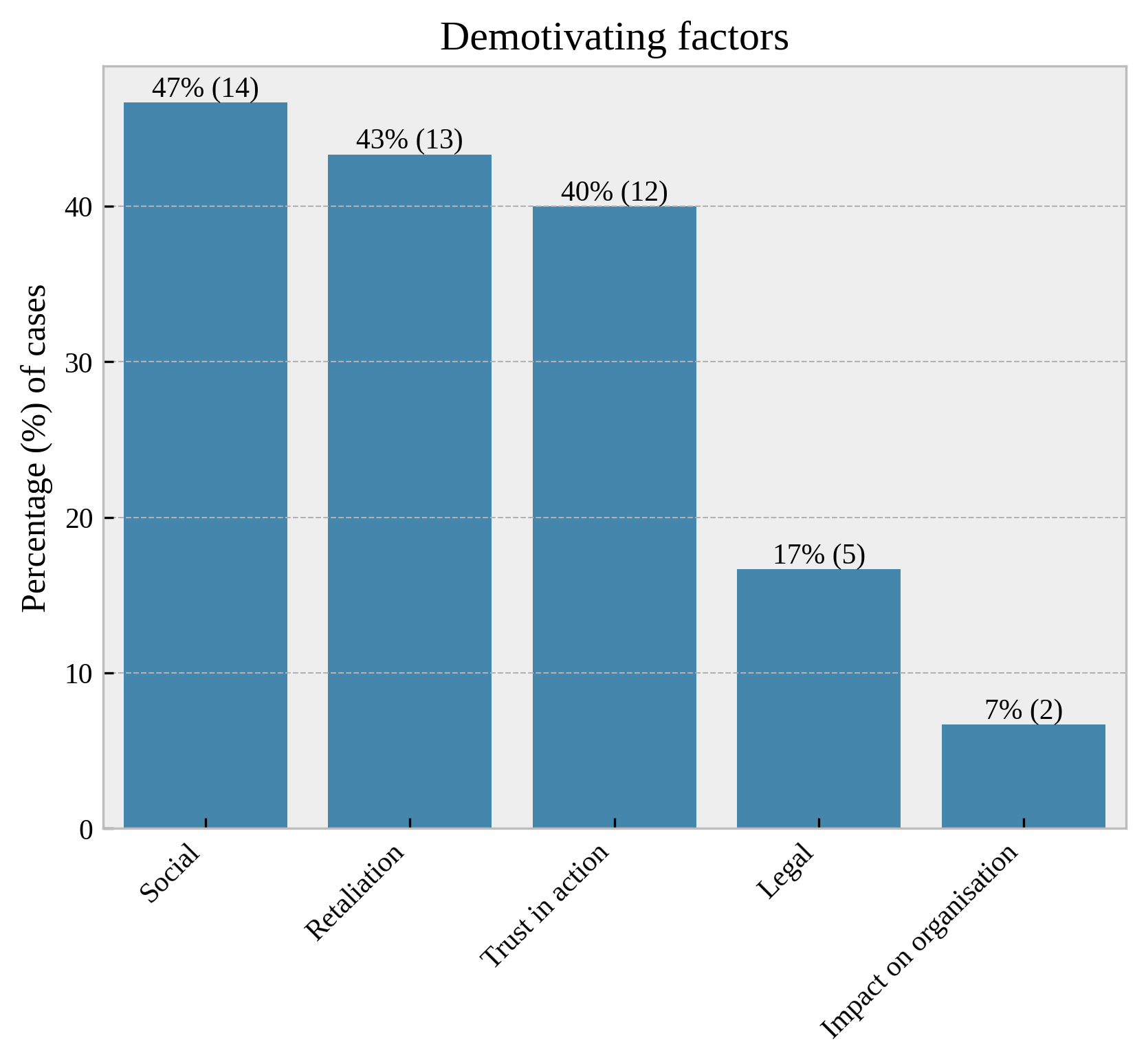}
    \caption{Demotivating factors present for whistleblowers. All percentages are lower bounds.}
\end{figure}

\begin{table}[!h]
    \caption{Demotivating factors, including "Unclear" cases.}
    \centering
    \begin{tabular}{cccccc}
        \toprule
        & Impact on organisation & Retaliation & Legal consequences & Social & Trust in action being taken \\
        \midrule
        Yes & 2 (7\%) & 13 (43\%) & 5 (17\%) & 14 (47\%) & 12 (40\%) \\
        No & 25 (83\%) & 5 (17\%) & 12 (40\%) & 7 (23\%) & 5 (17\%) \\
        Unclear & 3 (10\%) & 12 (40\%) & 13 (43\%) & 9 (30\%) & 13 (43\%) \\
        \bottomrule
    \end{tabular}
\end{table}

See footnotes for the definitions of the "impact on organisation" demotivating factor\footnote{The "impact on organisation" demotivating factor was present if a whistleblower was deterred by their expectation of the impact of whistleblowing on the wrongdoer organisation.}, and the "social" demotivating factor\footnote{"Social" demotivating factors were present if the whistleblower stood to lose social status by whistleblowing, such as by being in a cultural environment where whistleblowing is strongly discouraged.}.

\section{Discussion}

\subsection{Whistleblowers almost always appeared to be (at least partly) morally motivated ($\ge$87\% of cases)}
This is compatible with another study of whistleblowers (also mentioned in Limitations), which found that their top three motivations were integrity, justice, and altruism \citep{tnejom2010}. This is a surprising result, which should invite healthy scepticism. \textbf{One critique which we would challenge, however, is that this result is due to reporting and self-reporting biases:} that is, whistleblowers and reporters having an incentive to misrepresent their motives (specifically, to report altruistic motives instead of more “selfish” ones). As discussed in Methodology and dataset, we were constantly aware of this source of bias, and made a significant effort to mitigate it: we took a number of reported perspectives into account, compared whistleblowers’ incentives to those of other parties, and used a range of sources. We encourage readers to explore the motivation and evidence fields of our dataset to see this for themselves.

\subsection{Whistleblowers were almost always employees of the organisation which performed the wrongdoing ($\ge$90\% of cases)}
This result is also compatible with the SEC’s finding that most whistleblowers (62\%) were insiders \citep{secreport2024}. This is perhaps not surprising: for many kinds of wrongdoing, it may be that the only people in a position to know about it would be those “on the inside”. This could also explain why our sample contained so many executives (at least 33\%): presumably, they would have the most access to information of any employee. 

\subsection{Very few whistleblowers sought anonymity (13\%)}
This is surprising: anonymity can be a useful tool in avoiding retaliation, as the wrongdoer has to at least know who to retaliate against. However, there are a number of reasons why a whistleblower may not seek anonymity, which are discussed at length in previous versions of this paper. Briefly, these include fame, financial gain (such as monetising one’s image), protection (counterintuitively: making one’s identity public also makes it clearer when one has been retaliated against, possibly deterring retaliation), complicity (“clearing one’s name”), naivety, and psychological stress (which could negatively affect one’s decision-making abilities). \textbf{The best explanations for low rates of anonymity appear to be sampling bias, and psychological stress.} Regarding the former: given our sample was taken from high-profile, famous cases—and given anonymous cases are presumably less likely to be famous—our sample likely underrepresents the anonymous portion of the whistleblower population. Regarding the latter: our general impression is that whistleblowers were often highly distressed, which likely affected their decision-making abilities. Discovering wrongdoing, contemplating whether to report it, and (in many cases) being retaliated against are all highly stressful events. This is supported by another study of twenty-six pharmaceutical whistleblowers, of which thirteen (50\%) reported stress-related health problems, such as panic attacks, insomnia, generalised anxiety, and more \citep{tnejom2010}.

\subsection{Most whistleblowers were retaliated against (57-67\%)}
Often, this went beyond harassment (27-37\%) to unjust termination (20-30\%) or even death threats (13-23\%). This underscores the importance of whistleblower protections—both in terms of employment, and in terms of whistleblowers’ personal safety.

\section{Recommendations}

\textbf{Our ten concrete recommendations are numbered and bolded,} with explanations provided below.

\subsection{Financial rewards}

\textbf{
\begin{enumerate}[series=recs]
    \item Whistleblowers seem more likely to report if they receive a large percentage of the sanction amount, similar to the SEC and CFTC programmes.
    \item To disincentivise law-breaking, sanctions must be high enough to offset the expected gain from breaking the law.
\end{enumerate}
}

Financial rewards played a moderate role in our dataset, and were often paired with other motives: there is just one case where it is possible that financial rewards were the \textit{only} motive\footnote{This is Harry Markopolos, who reported the Madoff Ponzi scheme.}. This should not be taken as evidence against the importance of financial rewards, however: only four whistleblowers in our sample reported through \textit{de jure} processes which could have financially rewarded them\footnote{These whistleblowers are: John Barnett (filed for whistleblower retaliation), and Peter Rost, Walter Tamosaitis, and David Franklin (False Claims Act; only Tamosaitis and Franklin were successful).}. Indeed, one literature review of the experimental evidence finds that financial rewards make individuals more likely to report wrongdoing \citep{wallm2023}, as predicted by the microeconomic literature \citep{givat2016, givat2017}. Additionally, the success of the SEC whistleblower programme—which rewards whistleblowers with 10-30\% of the value of any sanction levied as a result of their tip—\textbf{would suggest that there is a strong case for financial rewards.}

\textbf{We now address two common concerns about whistleblower rewards.} The first is that although rewards create an incentive to submit true tips, they also create an incentive to submit \textit{false} tips. However, what matters for compliance is not the number of false reports submitted—existing whistleblower offices sift through tens of thousands of unhelpful tips every year\footnote{“In FY 2024, the Commission received approximately 24,980 TCRs [roughly, “tips”]”. Only 47 whistleblowers received an award \citep{secreport2024}.}—but their probability of succeeding\footnote{ The economic intuition here is that if false reports succeed sufficiently often, firms will (roughly speaking) assume that they will be fined regardless of whether they comply with regulation. This creates an incentive for noncompliance. See p. 56 of \cite{givat2016}.}. If a whistleblower office can keep false positives to a minimum, there should be minimal effect on compliance. 

Another concern is that financial rewards might “crowd out” whistleblowers’ moral motivation. This is especially salient in light of our findings, in which at least 87\% of whistleblowers appeared to be at least partly morally motivated. \cite{wallm2023} suggests that the experimental evidence on this question is conflicting. \textbf{Overall, however, the benefits of financial rewards seem to outweigh their costs.}

\textbf{How much should whistleblowers be rewarded?} At most, they should receive less than the amount of the sanction\footnote{The reward should be less than the amount of the sanction in order to prevent the whistleblower and wrongdoer from colluding. If the reward was greater than the sanction, the whistleblower and wrongdoer could collude to break the law, incur the sanction, collect the reward, and then split the excess reward between themselves.}. Givati finds that the optimal whistleblower reward is increasing in 1) the expected gain to the wrongdoer from breaking the law, and 2) the personal cost to the whistleblower from reporting wrongdoing \citep{givat2016}. AI is especially high-stakes, meaning that both (1) and (2) are especially high. (1) is high because individuals, firms, or even states could stand to benefit greatly from breaking any law which gives them an advantage in the development of AI. (2) is high because whistleblowers may be sacrificing lucrative careers (especially if they work for top AI developers\footnote{For example, Sam Altman (CEO of OpenAI) claimed that Meta had offered several of his staff \$100 million signing bonuses \citep{alt2025}.}), and because the high stakes of AI may give wrongdoers extra incentive to retaliate. Additionally, the AI industry is relatively tight-knit\footnote{For example, Aschenbrenner frequently mentions “SF-gossip” and the “SF rumour mill”. “SF already feels like a peculiar AI-researcher-college-town; probably this won’t be so different. It’ll be the same weirdly-small circle sweating the scaling curves during the day and hanging out over the weekend…” \citep{aschen2024}.}, which may also lead to high social costs for whistleblowers (which was a demotivating factor for 47-77\% of the whistleblowers in our sample). 

\textbf{If sanctions are appropriately high—and scale with the expected gain to wrongdoers from breaking the law—then whistleblowers could be rewarded with a percentage of the sanction amount, similar to the SEC and CFTC programmes.} These programmes reward whistleblowers with 10-30\%, but—due to the high-stakes nature of the AI industry, and the lucrative careers that some whistleblowers may be giving up—\textbf{we would suggest a higher award percentage for whistleblowers, unless the amount of the sanctions are enough to provide the additional incentive needed.}

\textbf{This recommendation comes with some nuance.} One is that eligibility for rewards should possibly not include all persons equally. For example, some parties may already have significant incentives to find and report wrongdoing, such as auditors. This may make our proposed incentives excessive, but only for those who already have significant incentives to report wrongdoing. A second point is a tactical note for policymakers. Introducing any one policy in an extreme manner (e.g. introducing very high sanctions, and correspondingly high whistleblower rewards) may make AI companies more averse to future regulations. For example, a very generous whistleblower programme may be used as justification for resisting mandatory or voluntary audits. A strategy of more slowly introducing whistleblower programmes—such as gradually expanding the scope of laws they apply to, for instance—may be the better approach, depending on the industry and regulatory context.

\subsection{Protection}

\textbf{
\begin{enumerate}[resume=recs]
    \item Prohibit employers from retaliating against whistleblowers, as in Dodd-Frank Section 922. Ideally, this could also be \textit{proactively} enforced.
    \item Offer whistleblowers witness protection in cases involving threats of violence.
\end{enumerate}
}

Retaliation plays a big role in disincentivising whistleblowers: 57-67\% of the whistleblowers in our sample were retaliated against, and the fear of retaliation was a demotivating factor in over 40\% of cases. As a result, protecting whistleblowers from retaliation could make them significantly more likely to report wrongdoing.

Another option would be to grant S visas to international whistleblowers. This could be used to detect noncompliance occurring outside of U.S. territory, such as violations of export controls. S visas are also useful due to the global spread of the AI industry, ranging from top labs in the U.S. to semiconductor foundries in Taiwan: potential whistleblowers may reside all over the world.

\subsection{Anonymity}

\textbf{
\begin{enumerate}[resume=recs]
    \item Whistleblowers should be able to submit tips anonymously. Tipping could be done through a lawyer, or via an anonymous online platform.
    \item Cybersecurity measures should be taken by the whistleblower office in order to protect whistleblowers’ identities. 
\end{enumerate}
}

Few whistleblowers in our sample sought anonymity (13\%). However, this seems to be down to a combination of sampling bias and psychological distress (see Discussion). Being anonymous may reduce the personal cost of whistleblowing, by reducing the likelihood of retaliation and preventing any negative career or social impacts. \textbf{Therefore, enabling whistleblowers to tip anonymously could encourage more tipping.}

\subsection{Processing tips}

\textbf{
\begin{enumerate}[resume=recs]
    \item The AI whistleblower office should be well-staffed and well-funded in order to process tips effectively. Previous whistleblower programmes have not had adequate resources.
\end{enumerate}
}

Penalising organisations on the basis of false tips creates an incentive for wrongdoing (discussed under Financial Rewards). However, failing to penalise on the basis of true tips disincentivises whistleblowing, by reducing the expected value of whistleblower rewards. \textbf{Processing tips well is therefore highly important for attracting future tips.} However, many argue that the SEC programme’s “tip-sifting” function is severely understaffed, receiving tens of thousands of tips per year for a team of only 30-50 employees \citep{platt2023}. The AI whistleblower office should improve on this aspect of the SEC programme.

\subsection{Messaging and advice}

\textbf{
\begin{enumerate}[resume=recs]
    \item The office should raise awareness for the whistleblower programme using press releases and mandatory training.
    \item The office should build trust by signalling its effectiveness. For example, government officials and industry leaders could publicly endorse the programme. Additionally, adequate funding and staffing would signal that whistleblowing is taken seriously.
    \item The office could include a whistleblower advice body to help potential whistleblowers navigate the whistleblowing process, and determine whether they have reasonable cause to believe that a violation has occurred.
\end{enumerate}
}

\textbf{Potential whistleblowers need to be aware of the office, and need to trust it enough to tip.} There are a number of options for raising awareness: press releases for general publicity, and perhaps mandatory training to inform industry insiders about the whistleblowing process. The office could also build trust by signalling its effectiveness. This could be done by having government officials and industry leaders endorse the programme, as well as the office being adequately staffed and funded. \textbf{Previously, the SEC and CFTC programmes had a slow start, possibly due to messaging and awareness\footnote{For example, note the dramatic increase in the number of tips received by the SEC on p. 5 of \citet{secreport2023}.}. An AI whistleblower office should learn from this.}

\textbf{The AI whistleblower office should also offer advice.} Whistleblowers are often experiencing significant psychological stress, which seems to affect their decision-making ability (see Discussion). Additionally, violations of AI regulations may not be obvious or clear-cut. For example, it may be difficult for a would-be whistleblower to determine whether they have reasonable cause to believe that they are developing a potentially high-risk system before it has been released and used. To help them, \textbf{the office could include a body which determines whether potential whistleblowers have reasonable cause.}

\section{Conclusion}
Whistleblower programmes could make a range of AI regulations more effective. Designing a whistleblower programme involves answering two questions: one, how to attract tips, and two, how to process them. We have focused on the former question, though it is worth noting that both are highly important and are closely interrelated. To answer this question, we focused on understanding what motivates whistleblowers. We took an empirical approach, creating a rich dataset of 30 case studies of high-profile cases. Our data shows a number of common themes, including the prevalence of moral motivation, that most whistleblowers were insiders, and that most whistleblowers were the victims of retaliation. Our findings suggest that whistleblower programmes will be more effective if they financially reward whistleblowers, provide protections for whistleblowers, enable whistleblowers to report anonymously, are adequately staffed and funded, and provide advice to potential whistleblowers. 

\section*{Acknowledgements}
We would like to thank Erich Grunewald, Karl Koch, Jakub Krys, and Lisa Soder for their helpful feedback and discussion. We would also like to thank the Supervised Programme for Alignment Research, which connected Ethan Beri and Mauricio Baker. This report does not necessarily represent the views of acknowledged individuals or organisations, and any remaining mistakes are our own.

\newpage{}

\bibliographystyle{plainnat}
\bibliography{references}  

\end{document}